\documentclass{iopart}
\usepackage{accents} 
\usepackage{amssymb}
\usepackage{amsthm}

\newcommand{\wt}{\widetilde}
\newcommand{\wh}{\widehat}
\newcommand{\wb}{\overline}

\renewcommand{\th}[1]{\wh{\wt{#1}}}
\newcommand{\hb}[1]{\wb{\wh{#1}}}
\newcommand{\bt}[1]{\wt{\wb{#1}}}
\newcommand{\thb}[1]{\wh{\wt{\wb{u}}}}

\newcommand{\gT}{\mathsf{T}}
\newcommand{\gS}{\mathsf{S}}
\newcommand{\gSS}{{\wb{\mathsf{S}}}}
\newcommand{\gx}{\mathsf{x}}
\newcommand{\gy}{\mathsf{y}}
\newcommand{\gz}{\mathsf{z}}
\newcommand{\subgroup}{\mathsf{<}}
\newcommand{\rank}{{\mathrm{rank}}}
\newcommand{\Span}{{\mathrm{span}}}
\newcommand{\kernel}{{\mathrm{kernel}}}

\newtheorem*{thm}{Theorem}
\newenvironment{pf}{\noindent\ignorespaces{\bf Proof.} }{\\$\square$\par\smallskip\smallskip\noindent\ignorespacesafterend}

\begin{document}
\paper{Linear quadrilateral lattice equations and multidimensional consistency}
\author{James Atkinson}
\address{Department of Mathematics and Statistics, La Trobe University, Victoria 3086, Australia}
\date{\today}
\begin{abstract}
It is shown that every scalar linear quadrilateral lattice equation lies within a family of similar equations, members of which are compatible between one another on a higher dimensional lattice.
There turn out to be two such families, a natural parametrisation is given for each.
\end{abstract}
\pacs{02.30.Ik}
\section{Introduction}
By integrability we mean structure in {\it non-linear} systems, however it is quite natural for a specific criteria for integrability to also admit {\it linear} examples.
From the point of view of classification it is desirable to exhibit these systems.
There is also the possibility that the linear examples provide a useful context for methods of the non-linear theory. 
Or that the non-linear theory can shed light on linear problems, a point of view which has been successfully exploited for linear PDEs by Fokas and co-authors, for example see \cite{fokas}.

Loosely speaking, the multidimensional consistency of an equation means identifying the equation as lying within a parametrised family of similar equations, members of which are compatible between one another on a higher dimensional lattice.
This property, which was first made explicit in \cite{nw} and later also in \cite{bs}, is sufficient for the integrability of scalar quadrilateral lattice equations and has led, through pioneering work of Adler Bobenko and Suris, to classification results in this area \cite{abs1,hie,abs2}.
In these works the classification problem is tackled through the associated classification of polynomials of degree one in four variables which are compatible when associated to the faces of a cube.
This approach to classification also picks out compatible systems which are not symmetric between the various faces of the cube \cite{abs2} (in \cite{abs1,hie} additional symmetry was imposed and the non-symmetric cases were therefore excluded, non-symmetric systems on the cube were further discussed in \cite{atk,xp}). 

However, a remarkably strong result was formulated and proved in \cite{abs2} by introducing a notion of non-degeneracy for the defining polynomial, such polynomials were termed `type-Q'.
It was shown that every such polynomial lies in a unique (parametrised) compatible family of similar type-Q polynomials.
The classification question for scalar quadrilateral lattice equations based on the multidimensional consistency is not fully answered for equations defined by polynomials which are not of type-Q.
Certainly there are non-trivial examples, the polynomial defining the lattice potential KdV equation \cite{we,hir} is itself {\it not} of type-Q.
All {\it linear} equations also lie outside this class and it is these equations that are studied in the present paper.

We proceed by asking for compatibility between generic autonomous scalar linear quadrilateral lattice equations on a three dimensional lattice, this yields an algebraic system which constrains the coefficients of the equations.
It turns out that this system is tractable by methods from linear algebra.
Using the appropriate transformation group and a re-parametrisation we then suggest a canonical form for the compatible systems which emerge.

\section{Imposing consistency on the cube}
\begin{figure}[ht]
\begin{center}
\begin{picture}(200,140)
\input{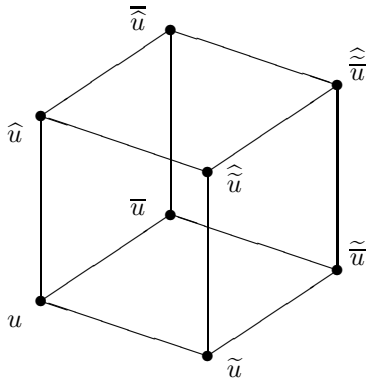}
\end{picture}
\end{center}
\caption{Variables associated to the vertices of a cube}
\label{cubepic}
\end{figure}

We will ask for compatibility between the following system of polynomial equations,
\begin{equation}
\fl \qquad
\begin{array}{cc}
a_1u+b_1\wt{u}+c_1\wh{u}+\th{u}=d_1,\qquad& a_1\wb{u}+b_1\bt{u}+c_1\hb{u}+\thb{u}=d_1,\\
a_2u+b_2\wh{u}+c_2\wb{u}+\hb{u}=d_2,\qquad& a_2\wt{u}+b_2\th{u}+c_2\bt{u}+\thb{u}=d_2,\\
a_3u+b_3\wb{u}+c_3\wt{u}+\bt{u}=d_3,\qquad& a_3\wh{u}+b_3\hb{u}+c_3\th{u}+\thb{u}=d_3.\\
\end{array}
\label{concon}
\end{equation}
Here $a_i,b_i,c_i\in \mathbb{C}\setminus \{0\}$ and $d_i\in\mathbb{C}$ are the {\it coefficients} of the polynomials.
By compatibility we mean that given initial data $u$, $\wt{u}$, $\wh{u}$ and $\wb{u}$ taken from $\mathbb{C}$, and after evaluation of the intermediate variables $\th{u}$, $\hb{u}$ and $\bt{u}$ using the equations on the left, the remaining equations on the right determine the {\it same} value for $\thb{u}$.
By assigning the variables involved to the vertices of a cube as in Figure \ref{cubepic} each equation in (\ref{concon}) may be associated to a face of the same cube.
This provides a convenient geometrical configuration to visualise this notion of consistency as well as giving the property its name \cite{nw,bs}.
The system of equations (\ref{concon}) is not the most general system of linear polynomial equations on the cube because equations associated to opposite faces coincide.
However, this restriction is natural because polynomial equations which are compatible in the sense described may be used to define an autonomous compatible three-dimensional lattice system (as in the next section).

It is a matter of straightforward calculation to verify that the system (\ref{concon}) is compatible as described if and only if the coefficients satisfy the following system of equations:
\begin{eqnarray}
\quad
\begin{array}{r}
b_1b_3-b_3c_2+c_2c_1=a_1,\\
b_2b_1-b_1c_3+c_3c_2=a_2,\\
b_3b_2-b_2c_1+c_1c_3=a_3,
\end{array}
\label{abc}
\\
\begin{array}{r}
(b_2\!-\!c_3)d_1 - (1+b_3)d_2 + (1+c_2)d_3 = 0,\\
(1+c_3)d_1+(b_3\!-\!c_1)d_2-(1+b_1)d_3 = 0,\\
-(1+b_2)d_1+(1+c_1)d_2+(b_1\!-\!c_2)d_3 = 0.
\end{array}
\label{hom}
\end{eqnarray}
In the remainder of this section we give a preliminary analysis of this system.
In particular we will separate out three cases based on the rank of the matrix emerging in (\ref{hom}) which we denote by $M$,
\begin{equation}
M:=
\left[\begin{array}{ccc}
b_2-c_3&-1-b_3&1+c_2\\
1+c_3&b_3-c_1&-1-b_1\\
-1-b_2&1+c_1&b_1-c_2
\end{array}\right].
\label{Mdef}
\end{equation}
There are only three cases because $\rank(M)\le 2$, so for example the last equation in (\ref{hom}) is redundant because it is a consequence of the first two.
Suppose first that the rank of $M$ is zero.
This happens only if all entries of $M$ vanish, from which the following is immediate,
\begin{equation}
\fl \qquad \rank(M)=0 \quad \Leftrightarrow \quad b_i=c_i=-1, \quad i\in\{1,2,3\}. 
\label{r0}
\end{equation}
Consider now the case that $\rank(M)=1$.
If $a_1$, $a_2$ and $a_3$ are fixed in terms of $b_i$, $c_i$, $i\in\{1,2,3\}$ by (\ref{abc}), then
\begin{equation}
\fl \qquad \rank(M)\le 1 \quad \Leftrightarrow \quad a_i+b_i+c_i+1=0, \quad i\in\{1,2,3\}.
\label{r1a}
\end{equation}
This is true because the condition $a_i+b_i+c_i+1=0$, $i\in\{1,2,3\}$ is equivalent to the vanishing of all cofactors of $M$ which is necessary and sufficient for $\rank(M)\le 1$.
The following is also true,
\begin{equation}
\fl \qquad \rank(M)=1 \quad \Rightarrow \quad \kernel(M)=\Span(v_1,v_2,v_3),\\
\label{r1b}
\end{equation}
where the vectors $v_1$, $v_2$ and $v_3$ are defined as
\begin{equation}
v_1:=\left[\begin{array}{c}1+b_1\\0\\1+c_3\end{array}\right],\quad
v_2:=\left[\begin{array}{c}1+c_1\\1+b_2\\0\end{array}\right],\quad
v_3:=\left[\begin{array}{c}0\\1+c_2\\1+b_3\end{array}\right].
\label{vdef}
\end{equation}
Assuming the condition on the right of (\ref{r1a}) holds it can be verified by calculation that the vectors $v_1$, $v_2$ and $v_3$ defined in (\ref{vdef}) lie in the kernel of $M$, and by further calculation that $\Span(v_1,v_2,v_3)$ has dimension less than 2 only if $b_i=c_i=-1$, $i\in\{1,2,3\}$.
But by (\ref{r0}) this latter condition violates $\rank(M)=1$, hence if $\rank(M)=1$ then $\Span(v_1,v_2,v_3)$ must have dimension 2, which completes the verification of (\ref{r1b}).
And finally,
\begin{equation}
\fl \qquad \rank(M)=2 \quad \Rightarrow \quad \kernel(M)=\Span(w),
\label{r2}
\end{equation}
where the vector $w$ is defined as
\begin{equation}
 \quad w:=\left[\begin{array}{c}a_1+b_1+c_1+1\\a_2+b_2+c_2+1\\a_3+b_3+c_3+1\end{array}\right],
\end{equation}
and again $a_1$, $a_2$ and $a_3$ are given in terms of $b_i$, $c_i$, $i\in\{1,2,3\}$ by (\ref{abc}).
A calculation shows that $w$ lies in the kernel of $M$, whilst by (\ref{r1a}) the vanishing of $w$ contradicts $\rank(M)=2$ so $\Span(w)$ has dimension 1 and (\ref{r2}) is verified.

\section{Suggested canonical forms}
We consider here the three dimensional lattice system
\begin{equation}
\begin{array}{c}
a_1u+b_1\wt{u}+c_1\wh{u}+\th{u}=d_1,\\
a_2u+b_2\wh{u}+c_2\wb{u}+\hb{u}=d_2,\\
a_3u+b_3\wb{u}+c_3\wt{u}+\bt{u}=d_3,
\end{array}
\label{sys}
\end{equation}
which is similar to the polynomial system (\ref{concon}), but where now $u=u(n,m,l)$, $\wt{u}=u(n+1,m,l)$, $\wh{u}=u(n,m+1,l)$, $\wb{u}=u(n,m,l+1)$ and $\th{u}=u(n+1,m+1,l)$ etc. are values of a scalar dependent variable $u$ as a function of three independent variables $n,m,l\in\mathbb{Z}$.
The coefficients $a_i,b_i,c_i\in\mathbb{C}\setminus\{0\}$ and $d_i\in\mathbb{C}$, $i\in\{1,2,3\}$ are assumed to be autonomous.
By the compatibility of (\ref{sys}) we will mean compatibility of the defining polynomials as described in the previous section, specifically that the coefficients satisfy the equations (\ref{abc}) and (\ref{hom}).

It is clear that the form (and compatibility) of (\ref{sys}) is preserved by the point transformation group $\gT$
\begin{equation}
\gT=\left\{\ u(n,m,l)\rightarrow \alpha u(n,m,l)+\beta \ | \ \alpha\in\mathbb{C}\setminus\{0\}, \beta\in\mathbb{C} \ \right\}.
\label{sinf}
\end{equation}
And if the coefficients are such that (\ref{sys}) is {\it invariant} under the action of an abelian subgroup $\gS\subgroup\gT$, then the form of (\ref{sys}) is also preserved by the transformation group $\gSS$,
\begin{equation}
\gSS=\left\{ \ u(n,m,l)\rightarrow [\gx^n\cdot\gy^m\cdot\gz^l](u(n,m,l)) \ | \ \gx,\gy,\gz \in \gS \ \right\},
\label{gg}
\end{equation}
elements of which are often referred to as {\it gauge} transformations.
(If (\ref{sys}) were not invariant under $\gS$, then transformations from $\gSS$ would render it non-autonomous.)
In the following theorem we use these transformations to suggest a canonical form for compatible systems (\ref{sys}).

\begin{thm}
By the action of point and gauge transformations, compatible systems {\rm (\ref{sys})} may be brought either to the form
\begin{equation}
\begin{array}{lll}
u-\wt{u}-\wh{u}+\th{u} = d_1,\\
u-\wh{u}-\wb{u}+\hb{u} = d_2,\\
u-\wb{u}-\wt{u}+\bt{u} = d_3,
\end{array}
\label{sys1}
\end{equation}
for some choice of the parameters $d_1,d_2,d_3\in\mathbb{C}$, or to the form
\begin{equation}
\begin{array}{ccc}
(p_1-q_1)u + (p_2-q_2)\th{u} = (p_2-q_1)\wt{u} + (p_1-q_2)\wh{u},\\
(q_1-r_1)u + (q_2-r_2)\hb{u} = (q_2-r_1)\wh{u} + (q_1-r_2)\wb{u},\\
(r_1-p_1)u + (r_2-p_2)\bt{u} = (r_2-p_1)\wb{u} + (r_1-p_2)\wt{u},
\end{array}
\label{sys2}
\end{equation}
for some choice of the parameters $p_i,q_i,r_i\in\mathbb{C}$, $i\in\{1,2\}$.
Conversely, either of the systems {\rm (\ref{sys1})} or {\rm (\ref{sys2})} are compatible for any choice of their respective parameters $d_1,d_2,d_3\in\mathbb{C}$ and $p_i,q_i,r_i\in\mathbb{C}$, $i\in\{1,2\}$.
\end{thm}
\begin{pf}
To the system (\ref{sys}) we associate a matrix $M$ defined by (\ref{Mdef}).
Compatible systems (\ref{sys}) for which $M$ is rank zero are necessarily of the form (\ref{sys1}), a fact which is evident by inspection of (\ref{r0}), (\ref{abc}) and (\ref{hom}).
We will now demonstrate that by point and gauge transformations, any compatible system (\ref{sys}) for which the associated matrix $M$ is of non-zero rank can be brought to a form where $\rank(M)\le 1$ and $d_1=d_2=d_3=0$ (we refer to systems satisfying this latter condition as homogeneous).

Consider first that $\rank(M)=1$.
Using (\ref{r1b}) we know that $d_1$, $d_2$ and $d_3$ are necessarily of the form
\begin{equation}
\begin{array}{l}
d_1 = \beta_1(1+b_1) + \beta_2(1+c_1),\\
d_2 = \beta_2(1+b_2) + \beta_3(1+c_2),\\
d_3 = \beta_3(1+b_3) + \beta_1(1+c_3),\\
\end{array}
\end{equation}
for some $\beta_1,\beta_2,\beta_3\in \mathbb{C}$.
But the condition on the coefficients appearing in (\ref{r1a}) is equivalent to the invariance of (\ref{sys}) under transformations of the form $u(n,m,l)\rightarrow u(n,m,l)+\beta$ for any $\beta\in\mathbb{C}$.
It follows that we may apply the gauge transformation 
\begin{equation}
u(n,m,l)\rightarrow u(n,m,l)+n\beta_1+m\beta_2+l\beta_3,
\end{equation}
which brings (\ref{sys}) to the homogeneous form.
Note that $M$ does not change under this transformation so that in particular we still have that $\rank(M)=1$.

Consider now the only other case, that $\rank(M)=2$.
It is immediate from (\ref{r2}) that $d_1$, $d_2$ and $d_3$ are necessarily of the form
\begin{equation}
\begin{array}{l}
d_1 = \delta(a_1+b_1+c_1+1),\\
d_2 = \delta(a_2+b_2+c_2+1),\\
d_3 = \delta(a_3+b_3+c_3+1),
\end{array}
\end{equation}
for some $\delta \in \mathbb{C}$.
Applying the point transformation $u(n,m,l)\rightarrow u(n,m,l)+\delta$ therefore brings (\ref{sys}) to the homogeneous form.
This transformation does not change $M$, however it does leave (\ref{sys}) with an invariance to point transformations of the form $u(n,m,l)\rightarrow \alpha u(n,m,l)$ for $\alpha\in\mathbb{C}\setminus\{0\}$.
We now claim that application of the gauge transformation
\begin{equation}
u(n,m,l) \rightarrow \alpha_1^n\alpha_2^m\alpha_3^l u(n,m,l),
\label{agauge}
\end{equation}
where the parameters $\alpha_1$, $\alpha_2$ and $\alpha_3$ are defined in terms of the coefficients and a new parameter $\kappa\in\mathbb{C}\setminus\{-1,1\}$ by the equations
\begin{eqnarray}
\alpha_1 = \frac{-2}{1+\kappa}c_1 + \frac{1-\kappa}{1+\kappa} b_3,\nonumber \\
\alpha_2 = \frac{1+\kappa}{1-\kappa} c_2 - \frac{2}{1-\kappa}b_1,\label{ai}\\
\alpha_3 = \frac{\kappa-1}{2} c_3 - \frac{1+\kappa}{2} b_2,\nonumber
\end{eqnarray}
brings the homogeneous case of (\ref{sys}) to a form in which the associated matrix $M$ is at most rank 1.
The claim can be verified by calculation, the calculation is equivalent to verifying that homogeneous systems (\ref{sys}) admit the solution $u(n,m,l)=\alpha_1^n\alpha_2^m\alpha_3^l u_0$ where $u_0\in\mathbb{C}$ is an arbitrary constant.
This follows because homogeneous systems (\ref{sys}) admit non-zero constant solutions only if the condition on the right of (\ref{r1a}) holds.
Note that to ensure invertibility of the transformation (\ref{agauge}) we should choose $\kappa\in\mathbb{C}\setminus\{-1,1\}$ so that none of $\alpha_1$, $\alpha_2$ or $\alpha_3$ vanish, inspecting (\ref{ai}) we see this can always be done, pathological cases are avoided because $b_i$, $c_i$, $i\in\{1,2,3\}$ are assumed non-zero.

So up to point and gauge transformations, we may assume that compatible systems (\ref{sys}) are of the form (\ref{sys1}) or else are homogeneous and satisfy the condition on the right of (\ref{r1a}).
It remains to show that systems of the latter form admit the parametrisation in (\ref{sys2}).

This final step in the proof is again through calculation, choose {\it distinct} parameters $p_2,q_2,r_2\in\mathbb{C}$ and define $p_1$, $q_1$ and $r_1$ by the equations
\begin{eqnarray}
p_1 = \frac{p_2-q_2}{q_2-r_2}r_2c_1 + \frac{r_2-p_2}{q_2-r_2}q_2b_3,\nonumber \\
q_1 = \frac{q_2-r_2}{r_2-p_2}p_2c_2 + \frac{p_2-q_2}{r_2-p_2}r_2b_1,\label{pqr}\\
r_1 = \frac{r_2-p_2}{p_2-q_2}q_2c_3 + \frac{q_2-r_2}{p_2-q_2}p_2b_2.\nonumber
\end{eqnarray}
Substitution of (\ref{pqr}) into (\ref{sys2}) followed by use of the condition on the right of (\ref{r1a}) and the relations (\ref{abc}) brings (\ref{sys2}) back to parametrisation in terms of the original coefficients.

The second statement of the theorem can be verified by calculation.
\end{pf}

The compatible system (\ref{sys2}) has been re-written in terms of {\it lattice parameters}, these are the two-component vectors $(p_1,p_2)$, $(q_1,q_2)$ and $(r_1,r_2)$.
By fixing an association between these three parameters the three lattice directions $n$, $m$ and $l$ respectively the system has an important property.
Namely {\it covariance}, it is invariant under permutations of the lattice directions.
The natural extension of the compatible system to any higher number of dimensions becomes apparent from this property.

We remark that the extension of (\ref{sys1}) to higher dimensions is also clear, but in this case it is natural do associate the constants $d_i$ each to a {\it pair} of lattice directions.
This has the consequence that the number of constants present in the system grows quadratically with the dimension of the system as opposed to linearly in the case of (\ref{sys2}).

We make a historical remark regarding the parametrisation of the compatible linear system (\ref{sys2}).
The parametrisation (\ref{sys2}) appeared in \cite{atk} in connection with a multidimensionally consistent equation discovered by Hietarinta \cite{hie1}.
The Hietarinta equation was shown to be linearisable originally in \cite{rjgt} with a different parametrisation given for the associated linear equation.
In the special case $p_2=-p_1$, $q_2=-q_1$ and $r_2=-r_1$ the system (\ref{sys2}) reduces to the equation found by linear approximation of the NQC equation \cite{nqc}, which is the equation satisfied by the discrete soliton plane-wave factors denoted $\rho_k$ in \cite{nqc}.
Note that the corresponding solution of (\ref{sys2}) is
\begin{equation}
u(n,m,l) = u_0 \left(\frac{p_1-k}{p_2-k}\right)^n\left(\frac{q_1-k}{q_2-k}\right)^m\left(\frac{r_1-k}{r_2-k}\right)^l.
\end{equation}

Finally we note that the two dimensional equation appearing in the homogeneous case of system (\ref{sys1}) was studied as an example in \cite{av}.

\section{Discussion}
It is easily verified that by the action of point and gauge transformations, quadrilateral lattice equations of the form
\begin{equation}
au+b\wt{u}+c\wh{u}+\th{u}=d,\label{leq}
\end{equation} 
where $a,b,c\in\mathbb{C}\setminus\{0\}$, $d\in\mathbb{C}$ and $u=u(n,m)$, $\wt{u}=u(n+1,m)$ etc., may be brought to one of the two forms
\begin{eqnarray}
u-\wt{u}-\wh{u}+\th{u}=d, &&\qquad d\in\mathbb{C}\label{c1},\\
u-\th{u}=\gamma(\wt{u}-\wh{u}), &&\qquad \gamma\in\mathbb{C}\setminus\{0\}.\label{c2}
\end{eqnarray}
Comparing this observation with the systems (\ref{sys1}) and (\ref{sys2}) occuring in the main result reveals that all scalar linear quadrilateral lattice equations are multidimensionally consistent.
But note that (\ref{c1}) has more potent solvability property beyond its linearity and multidimensional consistency, specifically for solutions of (\ref{c1}) it is true that
\begin{equation}
u(0,0)-u(n,0)-u(0,m)+u(n,m) = nmd,
\end{equation}
i.e., the equation can be directly integrated.
The principal example of the linear multidimensionally consistent equation is perhaps then the equation (\ref{c2}), and on comparing this with equations in the system (\ref{sys2}) there appears to be some redundancy in the three dimensional system.
However, choosing for example
\begin{equation}
p_1 = 1-\gamma, \quad q_1=-1-\gamma, \quad p_2=-1+\gamma, \quad q_2=1+\gamma,
\end{equation}
in (\ref{sys2}) removes the redundancy, although it breaks the covariance.
Actually the covariance is still present, but hidden:
Treating the system as an equation and its B\"acklund transformation (with B\"acklund parameter $(r_1,r_2)$) the covariant system emerges again as the superposition principle.
\ack
The author was supported by the Australian Research Council (ARC) Centre of Excellence for Mathematics and Statistics of Complex Systems (MASCOS).
He is very grateful to Frank Nijhoff, Maciej Nieszporski and Reinout Quispel for discussions.
He is also grateful for the hospitality of the Isaac Newton Institute for Mathematical Sciences (INI) in Cambridge where this article was completed during the programme Discrete Integrable Systems (DIS).

\end{document}